\documentclass[pra,twocolumn,showpacs]{revtex4-1}
\usepackage{amsmath}
\usepackage{graphicx}
\usepackage{fancyhdr}
\usepackage{float}
\usepackage{caption}
\usepackage{subcaption}

\begin{document}
\preprint{APS/123-QED}
\title{Quantum interference-enhanced deep sub-Doppler cooling of $^{39}$K atoms in gray molasses}
\author{Dipankar Nath}
\email{dip@tifr.res.in}
\affiliation{Tata Institute of Fundamental Research, Navy Nagar, Colaba, Mumbai-400005}
\author{R Kollengode Easwaran}
\affiliation{Tata Institute of Fundamental Research, Navy Nagar, Colaba, Mumbai-400005}
\author{G. Rajalakshmi}
\affiliation{Tata Institute of Fundamental Research, Navy Nagar, Colaba, Mumbai-400005}
\author{C.S. Unnikrishnan}
\email{unni@tifr.res.in}
\affiliation{Tata Institute of Fundamental Research, Navy Nagar, Colaba, Mumbai-400005}
\begin{abstract}
 We report enhanced sub-Doppler cooling of the bosonic atoms of $^{39}$K facilitated by formation of dark states tuned for the Raman resonance in the $\Lambda-$configuration near the D1 transition. Temperature of about 12 $\mu$K is achieved in the two stage D2-D1 molasses and spans a very large parameter region where quantum interference persists robustly. We also present results on enhanced radiation heating with sub-natural linewidth (0.07$\Gamma$) and signature Fano like profile of a coherently driven 3-level atomic system. The Optical Bloch Equations relevant for the three-level atom in bichromatic light field is solved with the method of continued fractions to show that cooling occurs only for a small velocity class of atoms, emphasizing the need for  pre-cooling in D2 molasses stage.
  
\end{abstract}

\maketitle
\section{Introduction}
Potassium and Lithium are important alkali group atoms for a variety of experiments employing cold atoms due to the existence of both fermionic and bosonic isotopes. However, laser cooling of their bosonic isotopes to the sub-Doppler temperatures is difficult due to the closely spaced hyperfine levels in the cooling transitions. Further, their evaporative cooling to degeneracy is complicated because of their unfavorable scattering properties. The latter issue is usually solved with either sympathetic cooling with other atoms like Rubidium \cite{GMO:2001,GR:2007} or using Feshbach resonances tuned with a magnetic field, in an optical trap \cite{CDE:2007}. To attain sub Doppler temperatures in $^{39}$K, multi-stage molasses cooling strategies with relatively large laser power and large detuning in the first stage have to be implemented, and yet, the lowest temperatures are almost 100 times larger than the recoil limit, compared to a factor of 10 in case of Rubidium. $^{39}$K has been cooled to temperatures as 
low as 30-40$\mu $K using D2 line molasses cooling \cite{CFO:1998,VGO:2011,MLA:2011}. 

Gray molasses aided by the formation of optically pumped dark state, with lasers tuned above the D1 line (blue detuning) of the alkali atoms have been shown to result in ultra-low sub-Doppler temperatures \cite{AHE:1995,DBO:1995}. Very recently, enhanced cooling due to quantum interference effects and coherent dark state formation in the Raman configuration involving two driving light modes and a 3-level atom have been reported in laser cooling of $^{40}$K \cite{DRF:2012} and $^{7}$Li \cite{ATG:2013}. Similar three level systems have also been used for velocity selective coherent population trapping (VSCPT) to cool atoms to sub-recoil temperatures \cite{AAS:1988}. In the case of Lithium, cooling below the Doppler limit is difficult without the use of special procedure, like cooling with the  narrow $^{2}S_{1/2} \rightarrow^3P_{3/2}$ transition in the UV \cite{PMD:2011}. While sub-Doppler cooling occurs naturally in $^{40}$K during the standard D2 molasses \cite{VGO:2011}, as its excited state hyperfine 
structure is well spaced, this is not true 
\begin{figure}[h]
 \includegraphics[width=.4\textwidth]{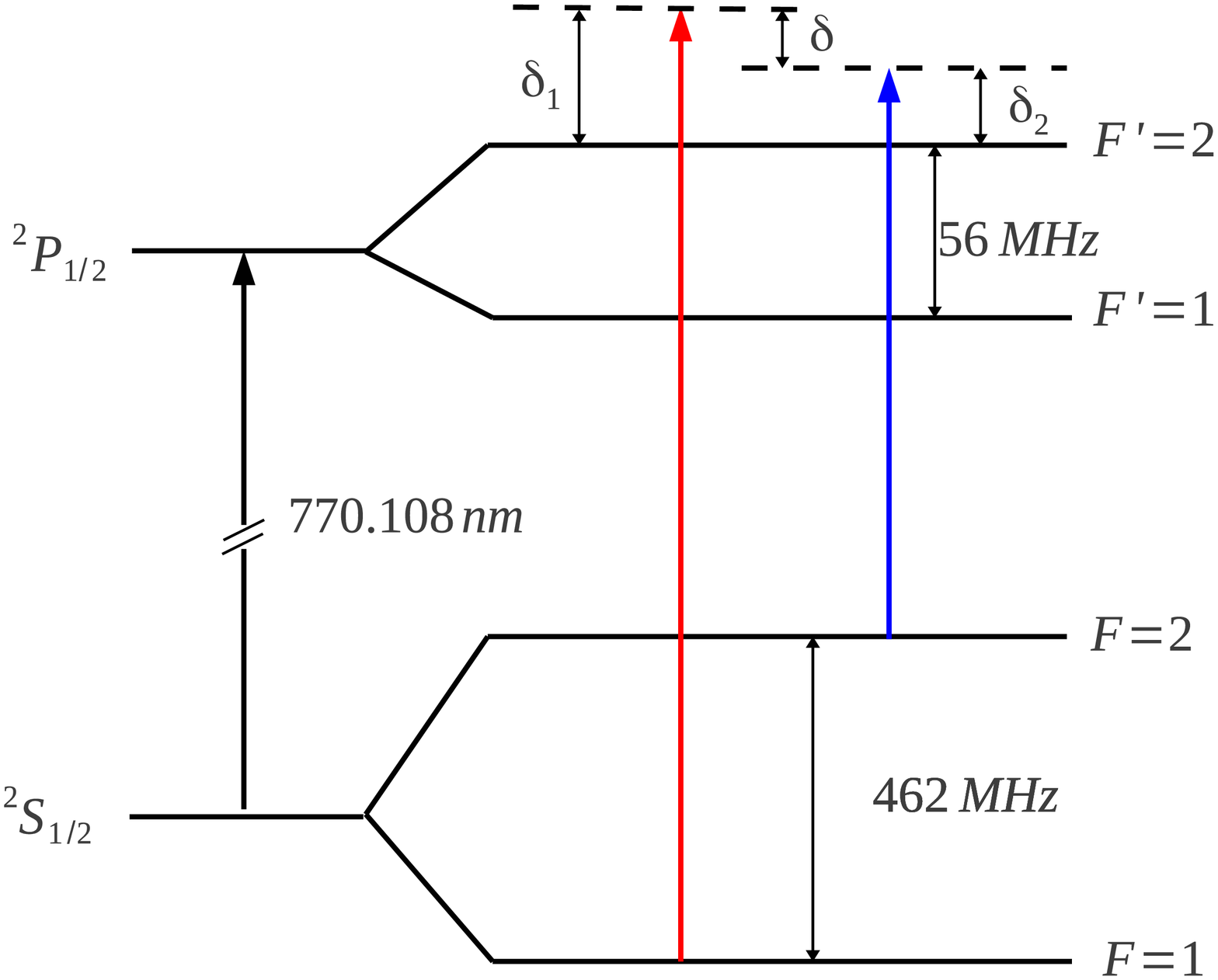}
 \caption{D1 line of $^{39}$K. $\delta_1, \delta_2$ are the repump ($F=1\rightarrow F'=2$) and cooling ($F=2\rightarrow F'=2$) detunings $\delta=\delta_1-\delta_2$ is the relative detuning between the repump and cooling beams. }
 \label{fig:D1line}
\end{figure}
for  $^{39}$K which requires special schemes for sub-doppler cooling \cite{ABA:1997,CFO:1998,VGO:2011,MLA:2011}.  We have been working on gray 
molasses with lasers tuned near the D1 line on the $^{39}$K atoms that are pre-cooled in a D2 conventional molasses (see Figure \ref{fig:D1line}) and observed that apart from optically pumped dark states aiding the sub-Doppler cooling by a factor of 2-3, quantum interference and coherent dark state formation helps to achieve sub-Doppler temperatures a further factor of 3 lower over a very wide detuning range, relative to both $F'=2$ and $F'=1$ D1 lines(See Figure \ref{fig:D1line}).
 While the exact Raman configuration always leads to enhanced cooling, there are regions of relative detuning of the cooling and repumping lasers where strong heating of the atomic cloud is observed. The heating follows an asymmetric Fano like profile with sub-natural linewidth ($< 0.1\Gamma$)  indicating coherence properties of the quantum superposition of the absorbing dressed state. The high reproducibility of the cooling and heating features and the vast parameter region over which the effects persist provides a new test system for the study of interplay between sub-Doppler laser cooling and quantum interference effects. 
In the rest of the paper, we report the main experimental results and their analysis, especially the ultra-low temperatures achieved in the molasses and the Fano-like profiles with sub-natural linewidth in radiation heating. We explain some of the observed phenomenon based on coherent population trapping effects in a three-level atom driven by two laser fields in $\Lambda$ configuration. A model for cooling mechanism in a bi-chromatic standing wave light field is developed by solving the optical Bloch equations by the method of continued fractions.

\section{Experiment and Results}
 The MOT is formed using 767nm laser tuned below the D2 transition of $^{39}$K atoms. Details of our experimental set up are described in \cite{VGO:2011}. The cooling and repump beams are derived from the same laser using AOMs and then mixed and amplified in a tapered amplifier. To capture the atoms into the MOT efficiently, the cooling beam is kept detuned by -24~MHz from $F=2\rightarrow F^{\prime }=3$ transition and the repump laser is detuned by -14~MHz from $F=1\rightarrow F^{\prime }=2$. We capture about $1\times10^{8} $ atoms in the MOT at a temperature of about 1~mK. The atoms are then put through a compressed-MOT (C-MOT) stage during which the magnetic field is ramped up and the cooling laser detuning is increased to -42~MHz. At the end of this stage the density of the trapped atoms is enhanced but the atoms are heated to temperatures $>$5~mK. Precooling of these atoms in a D2 sub-Doppler phase reduces their temperature to 40~$\mu $K \cite{VGO:2011}. The D2 molasses cooling lasts for about 4ms and 
during this phase the detuning of the cooling beam is reduced to -10~MHz and that of the repump beam to -11~MHz. Another laser (Toptica DLPro coupled to a home-made Tapered amplifier) tuned to 770.1 nm is used for
addressing the D1 transition. The `cooling laser' for the D1 $\Lambda $-molasses is near $F=2\rightarrow F^{\prime }=2$ transition with detuning $\delta_2$ and the `repump' is near \ $F=1\rightarrow F^{\prime }=2$ transition with detuning $\delta_1$. \ These are derived from the same beam using AOMs and the relative detuning $\delta =\delta_1-\delta_2$ as well as the absolute detuning ($\Delta$) from the D1 $F^{\prime }=2$ line are tunable over a wide range. The D1 laser beams are switched on immediately after the D2 sub-Doppler phase. This second D1 sub-Doppler phase cools the atoms by another factor of 4 or so in the optimal case of the parameter range we explored, down to about 12$\mu $K without loss of atoms. We achieve the lowest temperatures over a wide range of  blue detuning $\Delta $ of the cooling and the repump beams from $F=2\rightarrow F^{\prime }=2$ transition.

\begin{figure}[h]
 \centering
 \includegraphics[width=.5\textwidth]{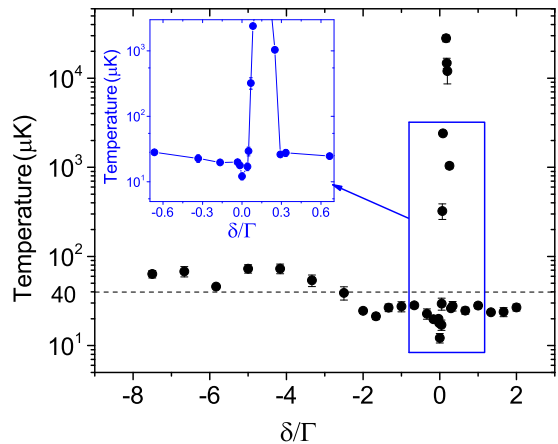}
 \caption{Temperature is measured for various values of the repump detuning, $\delta_1$, around the Raman condition for $\delta_2=$28MHz. The line at 40$\mu$K represents the starting temperature for the D1 sub-doppler.}
 \label{fig:tempvsrepumpdetuningat28MHz}
\end{figure}

Though we observe that cooling beyond the D2 molasses occur over a range of relative detuning $\delta ,$ extending $\pm 2\Gamma $ around $\delta =0,$
the deepest cooling occurs when the Raman resonance $\delta =0$ is satisfied. Figure \ref{fig:tempvsrepumpdetuningat28MHz} (with $\Delta =28$
MHz) indicates how the resonant cooling occurs in a very narrow band ($\ll\Gamma $) around the Raman condition. The intensity ratio between the
cooling ($I_{c}$) and repump ($I_{r}$) beams is about 3 for optimum cooling, which is much smaller than the ratio used in the case of $^{7}$Li \cite{ATG:2013}. For larger ratios of $I_c/I_r$ the cooling is less efficient. Temperature as low as 12\thinspace $\mu$K is observed at the Raman resonance.
For small positive values of the difference in the detuning $\delta =\delta_1-\delta_2$ we see strong heating. The heating is almost divergent by a factor of 1000 to temperatures beyond 10\thinspace mK and has a width of less than $0.1\Gamma $. The asymmetric Fano profile, typical of excitation probability in such atomic systems with interfering excitation pathways \cite{BLO:1992} is visible clearly in detailed measurements as shown in the inset of Figure \ref{fig:tempvsrepumpdetuningat28MHz}). 

\begin{figure}[h]
 \centering
  \includegraphics[width=0.5\textwidth]{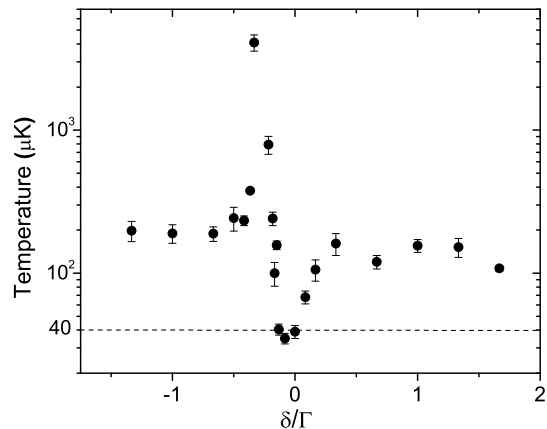}
  \caption{Plot for temperature vs $\delta$ for $I_C/I_R=0.28$ }
  \label{fig:tempvsrepumpdetuninginversedcoolingtorepumpratio}
\end{figure}

The two lasers driving the transitions in our experiments on D1 $\Lambda -$ dark state cooling have an intensity ratio, $I_{c}/I_{r}=\left(\Omega_2/\Omega _1\right) ^{2}\simeq 3.5$ when these parameters are optimized for best sub-Doppler cooling. This has significant implications on the exact nature of the dark state formation. If we invert the intensity ratio, i.e, make repump stronger than the cooling, the fano-profile shifts sides and appears at $\delta < 0$ (Figure  \ref{fig:tempvsrepumpdetuninginversedcoolingtorepumpratio}), confirming the role of coherent population trapping in the light scattering as explained in section \ref{sec:model}. Though the extra sub-Doppler cooling at the Raman resonance and the heating are not as effective here as in the case of $I_{c}/I_{r}\gg 1,$ the general 
trend is 
clear. Also, additional cooling happens only at the Raman resonance and at all other detuning we observe some mild heating, with the strong resonant heating to several mK occurring in a narrow range of $\delta<0,$ of width much smaller than $\Gamma $, close to the Raman resonance.

\begin{figure}[h]
  \centering
  \includegraphics[width=.5\textwidth]{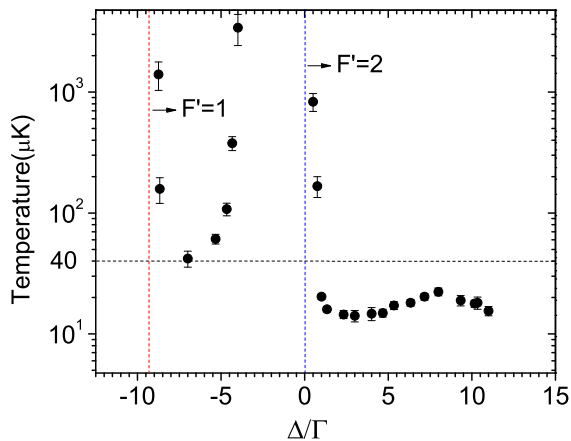}
  \caption{Temperature as a function of the detuning from $F'=2$ in the Raman configuration($\Delta=\delta_2=\delta_1$)}
  \label{fig:tempvsdetuning}
\end{figure}

The molasses cooling in Raman configuration is  effective over a wide range of absolute detuning ($\Delta =\delta_1=\delta_2$) from the atomic transition as can be seen from Figure \ref{fig:tempvsdetuning}. Over a range of $\Delta \approx (60,10)$ MHz, the quantum coherence-enhanced cooling is relatively independent of the value of $\Delta$. Strong heating takes over at $F^{\prime }=2$ resonance and
persists till $\Delta\approx-42$ MHz from $F^{\prime }=2$ line, below which we recover the temperatures reached in the D2 sub-Doppler cooling for a small range of detuning. When the lasers are tuned between $F^{\prime}=1$ and $F^{\prime }=2$ preserving the Raman resonance condition, two $\Lambda $-configurations become operational in the atom, namely the transitions ($F=1\rightarrow F^{\prime }=2,F=2\rightarrow F^{\prime }=2$) and ($F=1\rightarrow F^{\prime }=1,F=2\rightarrow F^{\prime }=1$). 
Clearly, the model for cooling or heating based on the 3-level $\Lambda $ configuration will not be adequate and the opposing contributions from the two
configurations largely explains the behavior in Figure \ref{fig:tempvsdetuning}. (See Section \ref{sec:model} for a discussion of the model)

\begin{figure}[h]
  \includegraphics[width=.5\textwidth]{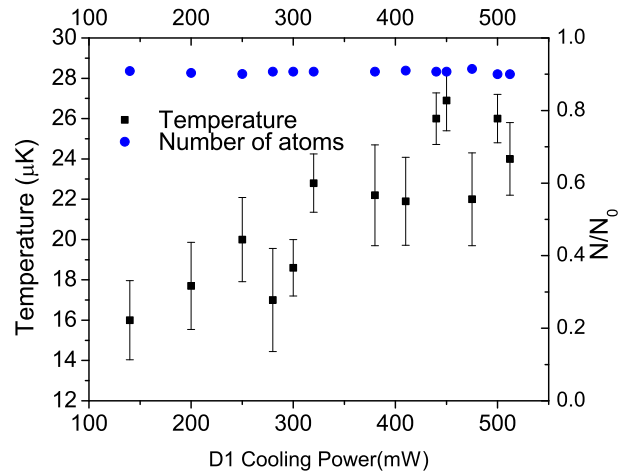}
  \caption{Temperature and fraction of atoms remaining in the cold cloud as function of cooling power of the D1 molasses. $N_0=10^8$ is the atom number prior to the molasses cooling stages }
  \label{fig:D1molasses-atom-loss}
 \end{figure}

 The loss of atoms during the D1 molasses stage has also been measured for various power in the cooling beam keeping the repump constant as shown in Figure \ref{fig:D1molasses-atom-loss}. About 90~\% of the MOT atoms remain after the D2-D1 molasses cooling stage.

\section{Model for Cooling}
\label{sec:model}
\begin{figure}[h]
 \includegraphics[width=0.5\textwidth]{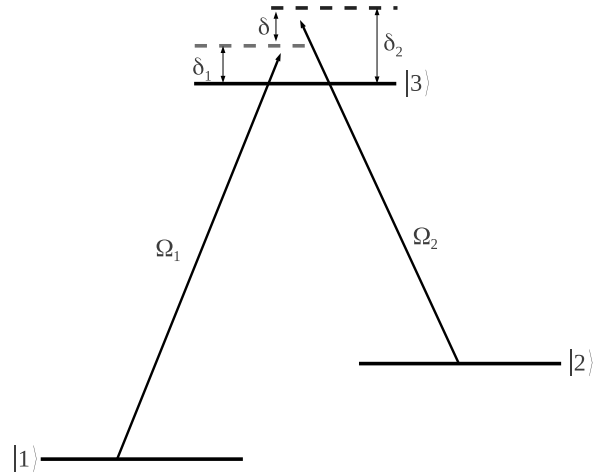}
 \caption{ A model three level $\Lambda-$system. $\Omega_1$ \& $\Omega_2$ are the Rabi frequencies of the transitions $|1\rangle\rightarrow|3\rangle$ \& $|2\rangle\rightarrow|3\rangle$ respectively. $\delta_1$ \& $\delta_2$ are the respective detuning and $\delta=\delta_1-\delta_2$ is the relative detuning. }
 \label{fig:model-three-level-system}
\end{figure}
The enhanced cooling and other main features in our results can be understood based on the absorption properties of a 3 level atomic system in the presence of $\Lambda$ type coherent drive fields. Referring to Figure \ref{fig:model-three-level-system}, the levels which correspond to the three level system in $\Lambda$ configuration are formed by the hyperfine ground state $F=1\left(|1\rangle\right)$ \& $F=2\left(|2\rangle\right)$  and the upper hyperfine D1 level $F'=2\left(|3\rangle\right)$. The transitions $|1\rangle\rightarrow|3\rangle$ and $|2\rangle\rightarrow|3\rangle$ are excited by laser light with frequencies $\omega_1$ \& $\omega_2$ respectively in standing wave configuration. The two waves with slightly different frequencies have a phase difference of $\phi$ between them.

Assuming that the two frequencies are almost similar i.e. $\omega_1\sim\omega_2=\omega$, we  can write the interaction part of the Hamiltonian of the combined atom-light system as shown in Equation \ref{eqn:approxhamiltonian}.

\begin{align}
H=&\hbar\Omega_1 \text{cos}\left(kz\right)\left(|1\rangle\langle3|\right)+h.c.&\nonumber\\&+\hbar\Omega_2 \text{cos}\left(kz+\phi\right)\left(|2\rangle\langle3|\right)+h.c.&\nonumber\\&+\hbar\delta_1|1\rangle\langle1|+\hbar\delta_2|2\rangle\langle2|&
 \label{eqn:approxhamiltonian}
\end{align}

  To get a qualitative model of the cooling process we consider a simplified picture where the one of the intensities is much larger than the other. We assume the repump intensity is relatively weak i.e. $\Omega_1\ll\Omega_2$. The new dressed ``bright'' states are given by \cite{CCT:1992}: 

\begin{align}
 |2'\rangle&=\text{cos}\theta\ |2\rangle-\text{sin}\theta\ |3\rangle&\nonumber\\
 |3'\rangle&=\text{sin}\theta\ |2\rangle+\text{cos}\theta\ |3\rangle&
 \label{eqn:dressedstates}
\end{align}
where $\text{tan}\ 2\theta=\Omega_2/\delta_2$. The line-width of both these ``mixed'' states are predominantly decided by the coefficient of $|3\rangle$ since the state $|2\rangle$ has an infinite lifetime and hence has a zero natural line-width. Thus the line-widths of the two new states are $\Gamma_{|2'\rangle}\sim\text{sin}^2\theta\ \Gamma\approx \left(\Omega_2/\delta_2\right)^2\Gamma$ and $\Gamma_{|3'\rangle}\sim\text{cos}^2\theta\ \Gamma\approx\Gamma$ (assuming $\Omega_2<\delta_2$). The state $|2'\rangle$ has a line-width narrower than $\Gamma$ by a factor $\left(\Omega_2/\delta_2\right)^2$, which give rise to the sub-natural line-width structures seen in the temperature vs relative detuning plot (Figure \ref{fig:tempvsrepumpdetuningat28MHz}). The other bright state $|3'\rangle$ has a line width similar to the line-width of the excited state $|3\rangle$.

Under the Raman condition $\delta _1=\delta _2$ coherent population trapping occurs in a $\Lambda $-system \cite{AAS:1988}. It is well known that the Raman condition leads to the formation of a ``dark'' state which do not couple to any other state. This non-coupling state is given by \cite{MWE:1994,MFL:2005}
\begin{equation}
 |NC\rangle=\left(\Omega_2 |1\rangle-\Omega_1 |2\rangle\right)/\left(\Omega_1^2+\Omega_2^2\right)
 \label{eqn:darkstate}
\end{equation}

The interaction of the two laser beams with the three level atoms thus results in three new perturbed energy levels and eigenstates.  The fluorescence of the system would now exhibit three resonances, if we scan the frequency of one of lasers, say the repump. For an atom moving in a standing wave light field, the light shift would be spatially modulated and would lead to a Sisyphus like 
cooling mechanism \cite{DRF:2012,ATG:2013}. The resonance in the light scattering would produce similar resonance in the cooling process which can be used to explain the features of Figure \ref{fig:tempvsrepumpdetuningat28MHz} and Figure \ref{fig:tempvsrepumpdetuninginversedcoolingtorepumpratio}. 

A more quantitative explanation for the observed cooling phenomena is provided by calculating the force on the atoms which can be obtained by solving the Optical Bloch Equations (OBEs) for the three level system. The optical field as seen by the atom can be written in the form given by Equation \ref{eqn:opticalfield} \cite{DVK:1994}.

\begin{equation}
 \vec{E}=\hat{e}[E_1\text{cos}\left(\omega_1t\right)\text{cos}\left(k_1z\right)+E_2\text{cos}\left(\omega_2t\right)\text{cos}\left(k_2z+\phi\right)]
 \label{eqn:opticalfield}
\end{equation}
where $\hat{e}$ is the polarization unit vector, $E_i, \omega_i$ and $k_i$ ($i=1,2$) are the electric field intensity, angular frequency resonant to the transitions $|i\rangle\rightarrow|3\rangle$ and corresponding wave vectors respectively. Here $\phi$ is the phase difference between the waves $\omega_1$ and $\omega_2$.

The force on the atoms (in one dimension) in the presence of a bi-chromatic beam in the case of a three level $\Lambda-$ system is given by:
\begin{align}
 F=&-\hbar k[\Omega_1\left(\rho_{13}+\rho_{31}\right)\text{sin}\left(kz\right)&\nonumber\\&+\Omega_2\left(\rho_{23}+\rho_{32}\right)\text{sin}\left(kz+\phi\right)]&
 \label{eqn:forceonatoms}
\end{align}
where we have assumed the wave-vectors to be equal i.e. $k_1=k_2=k$. $\rho_{ij}\left(i,j=1,2,3;i\neq j\right)$ are the coherence terms of the density matrix $\rho$.

To solve the OBEs we use the method of continued fractions developed in \cite{VSL:1981,DVK:1994}. The elements of the density matrix, $\rho_{ij}$, can be expressed in terms of its Fourier components (Equation \ref{eqn:fouriercomponents}).

\begin{equation}
 \rho_{lj} = \sum_{n=-\infty}^{n=+\infty}\rho_{lj}^{\left(n\right)} e^{inkz}
 \label{eqn:fouriercomponents}
\end{equation}

The average force on the atoms is determined by the first non-oscillating terms of the Fourier components and hence the average force on the atom computed over one wavelength can be written as (Equation \ref{eqn:averageforceonatoms}) \cite{DVK:1997},
\begin{align}
 &F=\frac{i \hbar k}{2}[\Omega_1\left(\rho_{31}^{\left(-1\right)}+\rho_{13}^{\left(-1\right)}-\rho_{31}^{\left(1\right)}-\rho_{13}^{\left(1\right)}\right)&\nonumber\\&+\Omega_2\left(\left(\rho_{32}^{\left(-1\right)}+\rho_{23}^{\left(-1\right)}\right)\exp{\left(i \phi\right)}-\left(\rho_{32}^{\left(1\right)}+\rho_{23}^{\left(1\right)}\right)\exp{\left(-i \phi\right)}\right)].&
 \label{eqn:averageforceonatoms}
\end{align}

\begin{figure}[h]
\includegraphics[width=0.45\textwidth]{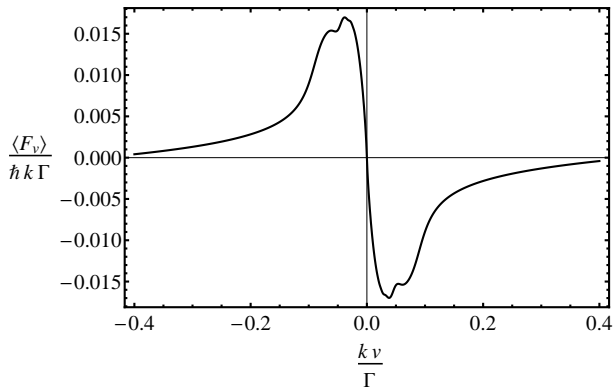}
\caption{Force on atoms averaged over phase, $\phi$  as function of velocity of the atoms for the parameters: $\Omega_1=0.7\Gamma$, $\Omega_2=1.2\Gamma$ and $\delta_1=\delta_2=5\Gamma$}
\label{fig:forceasafunctionofvelocity}
\end{figure}

Using the method of continued fractions we can now calculate the force on the atoms to arbitrary accuracy using parameters relevant to our experiment. Figure \ref{fig:forceasafunctionofvelocity} shows the force on an atom as a function of velocity. We calculate the force, taking into account that the relative phase ($\phi$) between the two light beams might undergo random changes on the time scales of ms.  The force is thus, calculated by averaging over the phase ($\phi$) between the two wavevectors $k_1$ and $k_2$ over the range ($0,2\pi$) \cite{ATG:2013}.
\begin{equation}
 \langle F_{v}\rangle=\frac{1}{2\pi}\displaystyle\int_0^{2\pi}F_vd\phi
 \label{eqn:phaseaveragedforce}
\end{equation}

The force, as we can see, produces effective cooling in a very narrow range of velocities in the Raman condition i.e. when $\delta_1=\delta_2=\delta$. Pre-cooling thus helps us in preventing any further loss of atoms during the D1 molasses cooling phase. Since the atoms in our experiment are already cooled down to about 40~$\mu K$ (corresponds to $kv/\Gamma \sim$0.02) prior to the D1 molasses stage, the loss of atoms is minimized. In fact we see that more than 90~\% of atoms are captured and cooled. Atoms with velocities significantly higher than the capture velocity of the force would undergo heating and would be lost.

\begin{figure}[h]
 \includegraphics[width=0.45\textwidth]{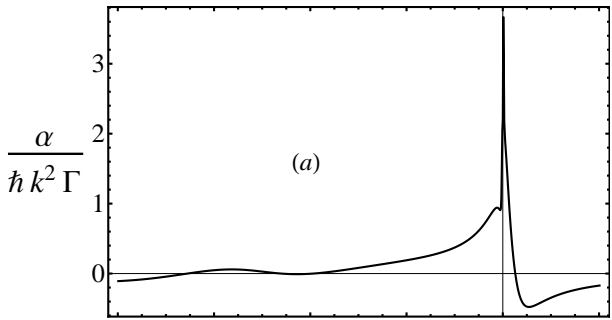}
 \caption{Coefficient of viscosity as a function of relative detuning ($\delta$). The parameters for the plot are $\Omega_1=0.7\gamma$, $\Omega_2=1.2\Gamma$, $\delta_2=5\Gamma$ and $\phi=0$}
 \label{fig:alphavsdetuning}
\end{figure}

From Figure \ref{fig:forceasafunctionofvelocity}, we can see that the force is linear for small velocities ($v$). The force ($F$) around $v=0$ can be written as
\begin{equation}
 F=-\alpha v
 \label{eqn:viscousforce}
\end{equation}
where $\alpha$ is the coefficient of viscosity. When $\alpha>0$, the force is opposite to the direction of velocity and we expect the atoms to cool and if $\alpha<0$ the force is in the same direction as that of velcity  and we expect heating to occur. In Figure \ref{fig:alphavsdetuning} we plot the coefficient of viscosity as a function of the relative detuning $\left(\delta=\delta_1-\delta_2\right)$ of the repump beam. The coefficient of viscosity ($\alpha$) is obtained from the linear fit of force around $v=0$. 
Figure \ref{fig:alphavsdetuning} shows the variation of coefficient of viscosity with relative detuning for the case $\Omega_2^2/\Omega_1^2 > 1$, i.e, $(I_C/I_R>1)$.  The cooling is maximum in the Raman condition (where $\alpha$ is large as well as $>0$) and as the relative detuning is increased it is seen that $\alpha<0$ and we expect to observe heating around this region. Thus as seen in Figure \ref{fig:tempvsrepumpdetuningat28MHz}, when we start moving from $\delta=0$(where the minimum temperature is attained) towards $\delta>0$, a large increase in temperature can be seen. For an inverted cooling to repump intensity i.e. $\Omega_2^2/\Omega_1^2 <1$ or $(I_C/I_R<1)$, as in the previous case the cooling is maximum in the Raman condition (when $\alpha>0$ and also maximum). However we expect heating to occur for negative values of the relative detuning ($\delta$) due to reversal of the symmetery. This explains the observed phenomena in Figure \ref{fig:tempvsrepumpdetuninginversedcoolingtorepumpratio}, where 
the heating takes place in negative side of the Raman condition i.e. in the region $\delta<0$.

\begin{figure}[h]
 \includegraphics[width=0.45\textwidth]{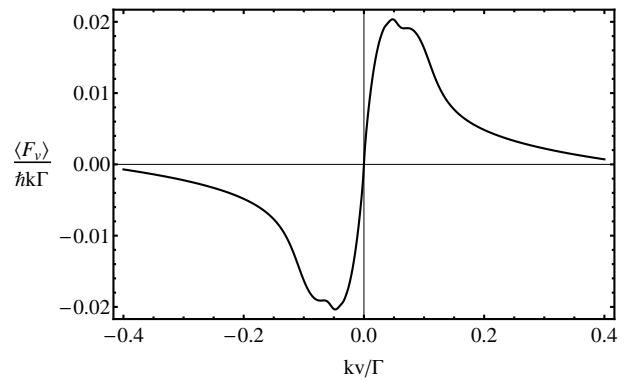}
 \caption{Force on atoms averaged over phase, $\phi$  as function of velocity of the atoms for the parameters: $\Omega_1=0.7\Gamma$, $\Omega_2=1.2\Gamma$ and $\delta_1=\delta_2=-4\Gamma$}
 \label{fig:forcevsvelocitynegativedetuning}
\end{figure}
In the model that we have chosen, the coefficient of  viscosity becomes negative when the detuning is negative i.e. the force is in the same direction as velocity and this should lead to heating of the cloud of atoms (See Figure \ref{fig:forcevsvelocitynegativedetuning}). This is confirmed by data for $\Delta<0$ relative to $F'=2$ as long as the magnitude of the detuning from $F'=1$ is much larger than the detuning from $F'=2$. As we go below the resonance of the $\Lambda-$ system under consideration, the other $\Lambda$ configuration that is present in our system, but hasn't been considered in our model, will start contributing to the total scattering. The second  $\Lambda-$system is formed by the transitions $F=1\rightarrow F'=1$ \& $F=2\rightarrow F'=1$. In order to understand the effect of the second $\Lambda-$ system, we treat both the $\Lambda-$systems as independent entities taking into account the relative transition strength of individual transitions in their respective Rabi frequencies. With this 
assumption, we can calculate 
the force on the atom due to the presence of each $\Lambda-$ system and then add the forces obtained individually. 

\begin{figure}[h]
 \includegraphics[width=0.5\textwidth]{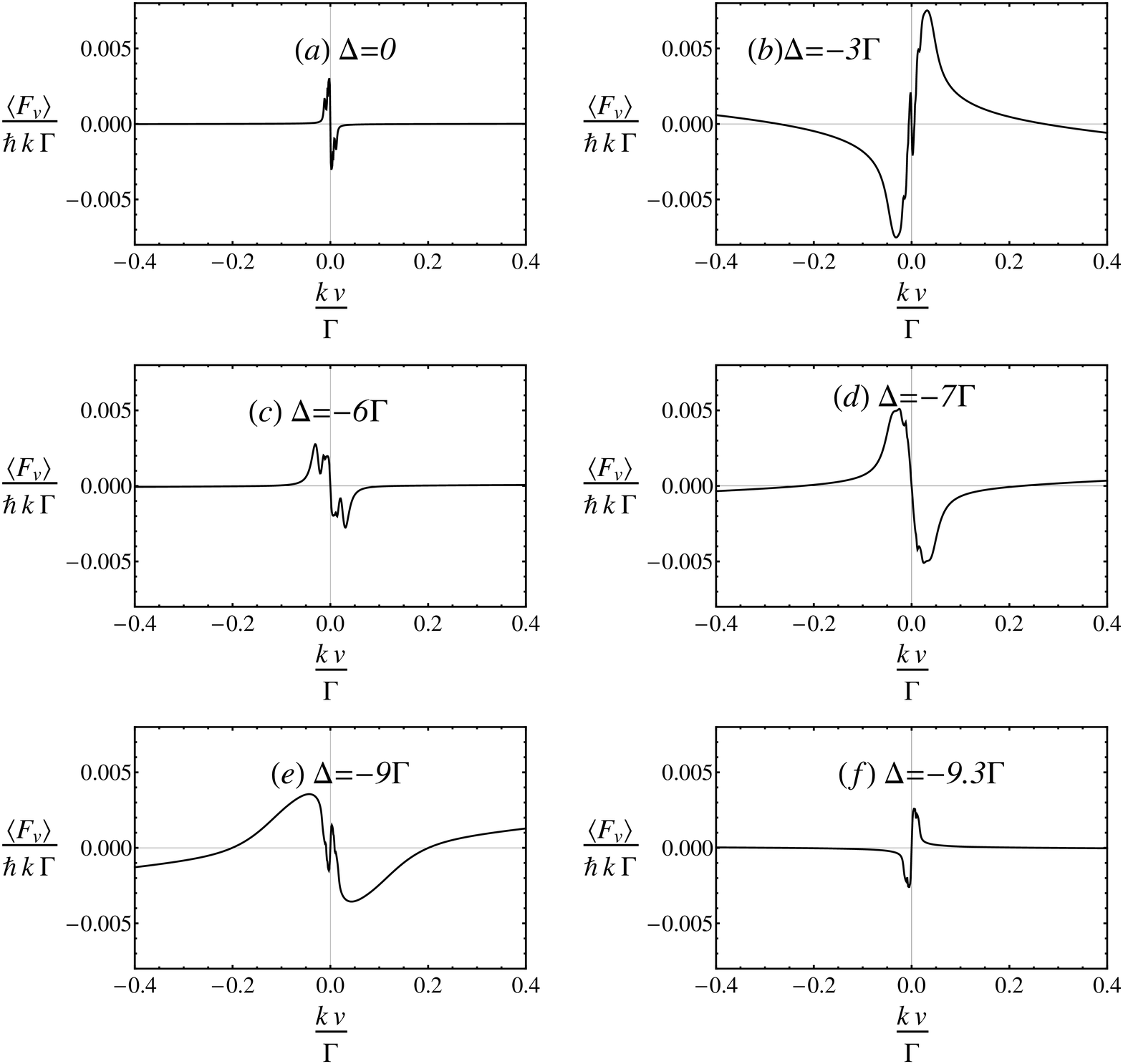}
 \caption{Force for several cases of absolute negative detuning ($\Delta$). Here $\Delta=\delta_1=\delta_2$. Plot ($a$) corresponds to $F'=2$ resonance and Plot ($f$) to $F'=1$ resonance.}
 \label{fig:forcevariousnegdetuning}
\end{figure}

Though simple addition of two forces is an approximation, the features seen can give us insights about the nature of the forces on the atom, when both the $\Lambda-$ systems actively participate in the cooling process. In a single absorption/emission cycle only one of the two $\Lambda-$systems present will participate and the force is calculated by averaging over one wavelength and the phase ($\phi$). Plot $\left(a\right)$ of Figure \ref{fig:forcevariousnegdetuning} corresponds to $\Delta=0$ where the contribution to the  force is entirely due to the presence of the second $\Lambda-$ system. When the laser frequency is red detuned by a small amount with respect $F'=2$ strong heating occurs as expected (see Figure \ref{fig:forcevariousnegdetuning}$\left(b\right)$).  As the laser frequency is 
further detuned heating dominates until $\Delta>-5\Gamma$. At about $\Delta=-6\Gamma$ (see Figure \ref{fig:forcevariousnegdetuning}$\left(c\right)$), the slope of the total force starts to become negative for a narrow velocity range leading to reduced heating. Blue Sisyphus cooling relative to $F'=1$ eventually overwhelms the heating from red detuning relative to $F'=2$ and we recover low temperature in the molasses for a small range of detuning. Closer to the $F'=1$ transition, the cooling reduces again for $\Delta<-9\Gamma$ before it completely converts to heating at the $F'=1$ resonance. Hence most features observed in figure \ref{fig:tempvsdetuning} is explained by our model involving two $\Lambda$-systems.

\section{Conclusion}
We have explored and established quantum coherent dark state enhanced cooling over a wide range of parameters. A plausible theoretical framework to analyze results on sub-Doppler cooling is presented which explains most of the observed features.  As the atoms move in the standing wave light field of the lasers detuned to the blue of the transition, the modulated light shift couples the atoms out of the dark state formed in the $\Lambda$ system. This leads to Sisyphus cooling for a narrow velocity class of atoms making it essential to pre-cool the atoms to temperatures below 100~$\mu$K to avoid atom loss. We achieve this by implementing a standard D2-molasses scheme prior to the molasses in the Raman configuration using the D1 levels. The two stage sub-Doppler cooling in molasses of $^{39}$~K results in a significant decrease in temperature to about 12~$\mu$K. Thus the atomic cloud is well-suited for transfer to magnetic or optical traps for further cooling to quantum degeneracy. 

\section{Acknowledgment}
 We thank Dr. Thomas Bourdel, Institut d' Optique, for discussions and for pointing out an important correction.

\bibliographystyle{apsrev}
\bibliography{D1SubDopplerCooling}

\end{document}